\documentclass[a4paper,10pt]{article}
\usepackage[utf8]{inputenc}
\usepackage{mathrsfs}
\usepackage{graphicx}
\usepackage{amsthm,amsfonts,amsmath} 
\newtheorem{prop}{Proposition}[section]
\newtheorem{lem}{Lemma}[section]

\newtheorem{thm}{Theorem}[section]

\newcommand{\p}{\partial}
\newcommand{\be}{\begin{equation}}
\newcommand{\ee}{\end{equation}}
\newcommand{\la}{\label}
\newcommand{\ba}{\begin{align}}
\newcommand{\ea}{\end{align}}

\begin{document}
\begin{center}
\section*{The Einstein equations and  multipole moments at null infinity}
\end{center}

-\centerline{J. Tafel}

\noindent
\centerline{Institute of Theoretical Physics, University of Warsaw,}
\centerline{Pasteura 5, 02-093 Warsaw, Poland, email: tafel@fuw.edu.pl}

\bigskip

\begin{abstract}
 We consider vacuum   metrics  admitting conformal compactification which is  smooth up to the scri $\mathscr{I^+}$. We write metric in the Bondi-Sachs form and expand it  into  power series in the inverse affine distance $1/r$. Like in the case of  the luminosity distance, given the news tensor and initial data for a part of metric the Einstein equations define coefficients of the series    in a  recursive way. This is also true in the stationary case however now the news tensor vanishes and the role of initial data is taken by multipole moments which are equivalent to moments of Thorne. We find an approximate form of metric and show that  in the case of vanishing mass  the mass dipole may be different from zero. Then the known result about the Kerr like behaviour of a stationary metric is violated. Finally we  find an approximate (up to the quadrupole moment) Bondi-Sachs form of the Kerr metric.
 
\end{abstract}

\bigskip

\null

\section{Introduction}
In 1958 Trautman \cite{t1,t2} began an era of theoretical description of gravitational waves in the full nonlinearized Einstein theory. He defined outgoing radiation conditions and showed that total energy cannot increase in time what was interpreted as an effect of the radiation. In 1960 Bondi \cite{b} presented his approach  based on a foliation of spacetime by null surfaces $u=const$ imitating null cones in the Minkowski space. Complete description of the axially symmetric case  was presented by Bondi, van der Burg and Metzner  \cite{bbm} and generalization to nonsymmetric metrics was given by Sachs \cite{s}. In this formalism  metric has a special form related to the null foliation. Metric coefficients are expanded into inverse powers of a radial coordinate $r$. The total energy  at $u=$const is defined as an integral of so called mass aspect which  generalizes the mass parameter in the Schwarzschild metric. The energy diminishes in time in a rate given by the square of the Bondi news function defined by low order metric coefficients. This fact is interpreted as an effect of gravitational radiation.

In the Bondi-Sachs approach one uses an intuitive notion of the null infinity available in the limit $r\rightarrow\infty$ when $u$ stays bounded. Its geometrical definition was  proposed by Penrose \cite{p} who considered spacetimes with metric $(\tilde M,\tilde g)$ admitting a conformal compactification to $(M,r^{-2}\tilde g)$ with a boundary containing the future null infinity $\mathscr{I^+}$.  Originally the conformal metric  was assumed to be smooth up to $\mathscr{I^+}$. In 1983 Friedrich \cite{f} and in 1985 Winicour \cite{w} noticed that this assumption may be too strong. Anderson and Chrusciel \cite {ac} suggested the so called polyhomogeneous expansion of metric admitting logarithmic terms  $r^{-n}\log^k{r}$. In 1995 Chruściel, MacCallum and Singleton \cite{cms} generalized the Bondi-Sachs formalism to  polyhomogeneous  expansions. Still the class of smooth conformal metrics is sufficiently big in many situations (see e.g. \cite{acf}).

In this paper we reexamine the vacuum Einstein equations  for metrics $\tilde g$ admitting smooth  scri $\mathscr{I^+}$.
We put  $\tilde g$ into the Bondi-Sachs form but, for geometrical reasons, we use the affine distance along null geodesics instead of the  luminosity one.
We expand the conformal metric and the Ricci tensor $\tilde R_{\mu\nu}$ into power series in $1/r$.
The low order Einstein equations have the most interesting consequences as  noticed by Bondi, Sachs  and others (see \cite{mw} for a review in the luminosity gauge). We present a method of  recursive solving of the equations  summarized in Theorem 2.1. This is not the existence theorem because to have this status one should prove convergence of the obtained series (note that for natural data for the considered situation  existence theorems are yet unknown). Still the recursive solution  can be useful for numerical computations (see \cite{w1} for known results and perspectives). In section 3 we consider stationary metrics. First we prove that these metrics undergo equations from section 2 (this is not a trivial observation since these equations are obtained for a specific choice of the conformal factor). From the low order equations we obtain  restrictions  which, for nonvanishing mass $M$, allow to write $\tilde g$ as an approximate Kerr metric. Solutions of higher order equations are given up to multipole moments which are equivalent to those of Thorne \cite {th} and Geroch \cite{g} and Hansen \cite{h}. The main results are summarized in Theorem 3.1. As an example we find approximate Bondi-Sachs coordinates for the Kerr metric and we write this metric up to terms defined by quadrupole moments.
%\end{document}

\section{The Einstein equations near conformal boundary}
In this section we reexamine the Einstein equations at null infinity in spirit of the Bondi-Sachs formalism combined with the Penrose conformal approach. The main difference between our results and those obtained by many authors (see \cite{mw} and references therein) is that our calculations are performed in the affine gauge. Theorem 2.1 should be considered not as a completely new  result but rather as a way to systemize recursive solving of the Einstein equations under assumption that solution with smooth scri does exist.

Following the Penrose compactification method 
we assume that spacetime $\tilde M$ with  metric $\tilde g$ can be partially compactified in a conformal way to $(M,g)$ with a future boundary $\mathscr{I^+}$ which can be foliated by  surfaces diffeomorphic to the 2-dimensional sphere, $\mathscr{I^+}=R\times \tilde S_2$. We assume that metric $g$ is smooth  in a neighbourhood $U$ of $\mathscr{I^+}$. In $U$ we introduce a system of the Bondi-Sachs coordinates in the following way.  First we define coordinates 
$u,x^A$ (with $A=2,3$) on $\mathscr{I^+}$ such that   $u=const$ on leaves of the foliation and vector field $\p_u$ is orthogonal to the foliation. Now from each point $p\in \mathscr{I^+}$ we emit a null geodesic with the tangent vector  $v$ orthogonal to the  foliation  and such that $g(v,\p_u)=-1$. We propagate coordinates $u,x^A$ along these geodesics and choose the fourth coordinate $\Omega$ to be the affine parameter along geodesics such that $\Omega=0$ and $\p_{\Omega}=v$ on $\mathscr{I^+}$.
In the coordinates $x^0=u$, $x^1=\Omega$ and $x^A$ the compactified metric takes the form
\be\la{1}
g=du(g_{00}du-2d\Omega+2 g_{0A}dx^A)+g_{AB}dx^Adx^B\ ,
\ee
where
\be\la{2}
\hat g_{0A} =0
\ee
(the hat  denotes value on $\mathscr{I^+}$).

Physical metric $\tilde g$  is related to $g$ via a conformal factor $\Omega'=f\Omega$, where $f$ is a function  nonvanishing and regular up to the boundary $\mathscr{I^+}$. In the coordinates $u,x^A$ and $r=\Omega^{-1}$ it is given by
\be\la{3}
\tilde g=du(\tilde g_{00}du+2f^{-2}dr+2\tilde g_{0A}dx^A)+\tilde g_{AB}dx^Adx^B\ ,
\ee
where $\tilde g_{00},\tilde g_{AB}$ are of the order $r^2$ and $\tilde g_{0A}=O(r)$. Taking coordinate $r'=\int{f^{-2}dr}$ instead of $r$ leads to elimination of  $f$
\be\la{4}
\tilde g=du(\tilde g'_{00}du+2dr'+2\tilde g'_{0A}dx^A)+\tilde g_{AB}dx^Adx^B\ .
\ee
Now, as  new compactified metric we  take $r'^{-2}\tilde g$, which  has the form (\ref{1}) with  $1/r'$ as the new coordinate $\Omega$. Thus, without a loss of generality we can assume that unphysical metric is (\ref{1}) and the physical metric is given by
\be\la{5}
\tilde g=\Omega^{-2} g\ .
\ee
 
It follows from (\ref{5}) that  the Einstein vacuum equations
 \be\la{6}
 \tilde R_{\mu\nu}=0
 \ee
 can be written in the form 
 \be\la{7}
  R_{\mu\nu}-2Y_{\mu\nu}-Yg_{\mu\nu}=0\ ,
 \ee
 where
 \be\la{8}
Y_{\mu\nu}=-\frac{1}{\Omega}\Omega_{|\mu\nu}+\frac {1}{2\Omega^{2}}\Omega_{|\alpha}\Omega^{|\alpha}g_{\mu\nu}\ ,\ \ Y=Y^{\ \alpha}_{\alpha}
 \ee
 and ${}_{|\mu}$ denotes the covariant derivative related to $g$. In the Bondi-Sachs coordinates nonvanishing components of  $g^{\mu\nu}$ are given by
 \be\la{9}
 g^{01}=-1\ ,\ \ g^{11}=-g_{00}+g_{0A}g_0^{\ A}\ ,\ \ g^{1A}=g_0^{\ A}\ ,\ \ g^{AB}\ ,
 \ee
 where $g_0^{\ A}=g^{AB}g_{0B}$ and $g^{AB}$ is an inverse matrix to  $g_{AB}$. Tensor $Y_{\mu\nu}$ and its trace take the form 
 \be\la{9a}
 Y_{\mu\nu}=\frac{1}{\Omega}\Gamma^{1}_{\ \mu\nu}+\frac {1}{2\Omega^2}g^{11}g_{\mu\nu}\ ,
 \ee
 \be\la{14}
Y=-\frac{1}{\Omega\sqrt{|g|}}(\sqrt{|g|}g^{1\alpha})_{,\alpha}+\frac {2}{\Omega^2}g^{11}\ ,
\ee
 where $\Gamma's$ denote the Christoffel symbols and
 \be\la{9b}
 |g|=\det{g_{AB}}\ .
 \ee
 
 We will expand metric  $g_{\mu\nu}$ into the Taylor series in $\Omega$ and study equations (\ref{7}) in all orders $\Omega^k$. Even if they can be solved 
 the resulting series does not have to be convergent. Nevertheless a cutoff of this series  approximates a true solution of the Einstein equations.
 We begin with a rather mild assumption that   $\tilde R_{\mu\nu}$ is finite on the boundary $\mathscr{I^+}$. Then  $Y_{\mu\nu}$ must be also finite. In order to avoid a second order pole at  $\Omega=0$ one has to assume the following expansion  of $g_{00}$ 
 \be\la{10}
g_{00}=a\Omega+b\Omega^2-2M\Omega^3+ O(\Omega^4)\ .
\ee
Now, components $Y_{1A}$ have no first order poles provided
\be\la{10a}
g_{0A}=q_A\Omega^2+2L_A\Omega^3+O(\Omega^4)
\ee
and regularity of $Y_{AB}$ is equivalent to
\be\la{11}
\hat g_{AB,0}=a\hat g_{AB}\ .
\ee
All remaining components of  $Y_{\mu\nu}$ are nonsingular at $\Omega=0$ under conditions (\ref{10})-(\ref{11}).

Equation (\ref{11}) implies that $\hat g_{AB}$ is proportional to an u-independent 2-dimensional metric which, thanks to the uniformization theorem and freedom of transformation of coordinates $x^A$, is proportional to the standard metric $s_{AB}$ of the 2-dimensional sphere $S_2$. Thus,
\be\la{11a}
\hat g_{AB}=-\gamma^2s_{AB}\ ,\ a=2(\ln{\gamma})_{,0}\ ,\ \gamma>0\ ,
\ee
where  $\gamma$ is a function of all variables. Now, we can choose new coordinates $\Omega'$ and $u'$ such that 
\be\la{11b}
\frac{\Omega'}{\Omega}\hat =\gamma\ ,\ \ u'\hat =\int{\gamma du}\ .
\ee 
This transformation leads to $\gamma'=1$ and $a'=0$. Hence, we can assume without  loss of generality that 
 \be\la{10b}
g_{00}=b\Omega^2-2M\Omega^3+ O(\Omega^4)
\ee
and
\be\la{12}
g_{AB}=-s_{AB}+n_{AB}\Omega +p_{AB}\Omega^2+O(\Omega^3)\ .
\ee
Still metric $s_{AB}$ is  defined up to the conformal group of the sphere, which together with ``supertranslations`` of $u$ form the Bondi-Metzner-Sachs group  of asymptotic symmetries. These transformations  can be also combined with a shift of $r$
\be\la{10c}
r'=r+h(u,x^A)+O(\Omega)
\ee
which can be used e. g. to obtain
\be\la{11d}
n=0\ ,
\ee
where 
\be
n=s^{AB}n_{AB}\ .
\ee

Before we start a more advanced analysis of the Einstein equations we will reduce their number by means of the Bianchi identity 
\be\la{36}
\tilde\nabla_{\mu} \tilde G^{\mu}_{\ \nu}=0
\ee
(we do it in a slightly different way than that of Bondi and Sachs, see \cite{w} for a comparison). To this end we write the  identity
in the form
\be\la{37}
2\Omega^2(\sqrt{|g|}\Omega^{-4}\tilde G^{\alpha}_{\ \nu})_{,\alpha}+\sqrt{|g|}g^{\alpha\beta}_{\ \ ,\nu}\tilde R_{\alpha\beta}+(\sqrt{|g|}\Omega^{-2})_{,\nu}\tilde R=0\ .
\ee
Note that $\alpha,\beta\neq 0$ in the middle term and 
\ba\la{38}
\Omega^{-2}\tilde R&=-2\tilde R_{01}+g^{11}\tilde R_{11}+2g^{1A}\tilde R_{1A}+g^{AB}\tilde R_{AB}\\\nonumber
\tilde G_{01}&=\frac 12(g^{11}\tilde R_{11}+2g^{1A}\tilde R_{1A}+g^{AB}\tilde R_{AB})\\
\tilde G_{1 1}&=\tilde R_{1 1}\ ,\ \ \tilde G_{A 1}=\tilde R_{1A }\ .\nonumber
\end{align}
In equation (\ref{37}) with $\nu=1$  function $\tilde R_{01}$ appears only in the last term. 
One obtains
\be\la{39}
\tilde R_{01}=\tilde G_{01}+f((\sqrt{|g|}\Omega^{-4}\tilde G^{\alpha}_{\ 1})_{,\alpha}+\frac 12\Omega^{-2}\sqrt{|g|}g^{\alpha\beta}_{\ \ ,1}\tilde R_{\alpha\beta})
\ee
provided that
\be\la{40}
f=((\sqrt{|g|}\Omega^{-2})_{,\Omega})^{-1}
\ee
is well defined. In a neighbourhood of the scri there is 
\be\la{41}
f=-\frac{\Omega^3}{2\sqrt{|\hat g|}}(1+O(\Omega))
\ee
  and from (\ref{39}) it follows that fullfilement of equations  
  \be\la{40a}
  \tilde R_{11}^{(l)}=\tilde R_{1A}^{(l)}=\tilde R_{AB}^{(l)}=0\ ,\ \ l\leq k
  \ee
  guarantees   $\tilde R^{(k)}_{01}=0$, where  $(k)$ denotes the k-th coefficient in the Taylor expansion in $\Omega$.
  If $\nu=A$ functions $\tilde R_{0\mu}$ appear in  (\ref{37}) only in the expression $-2\Omega^2(\sqrt{|g|}\Omega^{-2}\tilde R_{0A})_{,1}$ which vanishes if (\ref{40a}) is satisfied. Hence
   $\tilde R^{(k)}_{0A}=0$ for all $k$ except $k=2$. We  obtain  a similar result taking $\nu=0$. Thus, in order to solve the Einstein equations up to the order  $k\geq 2$  it is sufficient to consider (\ref{40a}) and 
   \be\la{40b}
   \tilde R^{(2)}_{00}=\tilde R^{(2)}_{0A}=0\ .
   \ee
  
 The simplest one from this reduced set of equations is   $\tilde R_{11}=0$. It 
 reads
  \be\la{a1}
  -\frac 12 (\ln{|g|})_{,11}+\frac 14g_{AB,1}g^{AB}_{\ \ ,1}=0\ .
  \ee
 In the order $k\geq 1$ it takes the form
\be\la{a2}
 \frac 12(k+1)(k+2)s^{AB}g^{(k+2)}_{AB}+\frac 12(k+1)^2n^{AB}g^{(k+1)}_{AB}=\langle g^{(l)}_{AB},l\leq k\rangle,\ k\geq 1\ ,
\ee
where  indicies  $A,B$ in $n^{AB}$ are raised by means of $s^{AB}$ and $\langle ...\rangle$ denotes an expression depending on variables in the bracket. For $k=0$ equation (\ref{a1}) yields
\be\la{a3}
p=-\frac 14n_{AB}n^{AB}\ ,
\ee
where $p=p^A_{\ A}$.
Thus, for all values  $k\geq 0$ one obtains
\be\la{a2a}
 s^{AB}g^{(k+2)}_{AB}=\langle g^{(l)}_{AB},l\leq k+1\rangle\ ,\ k\geq 0\ .
\ee

Equations $\tilde R_{1A}=0$  in the order $k=0$ define
\be\la{a5}
q_A=\frac 12n^B_{\ A|B}-\frac 12n_{,A}\ ,
\ee
where symbol $|A$ denotes the covariant derivative with respect to $s_{AB}$. 
Taking into account  (\ref{a1})  for $k\geq 1$  one obtains
\ba\la{a4}
\tilde R_{1A}^{(k)}-\frac 1k \tilde R_{11,A}^{(k-1)}=-\frac 12(k-1)(k+2)g^{(k+2)}_{0A}-\frac 12(k-1)n_A^{\ B}g_{0B}^{(k+1)}\\\nonumber\frac n4(k+1)g_{0A}^{(k+1)}
-\frac 12(k+1)(g_{AB}^{(k+1)})^{|B}+\langle g^{(l)}_{\mu\nu},l\leq k\rangle=0\ ,\ \ k\geq 1\ .
\end{align}
For $k\neq 1$ it follows from (\ref{a5}) and (\ref{a4}) that 
\ba\la{a4a}
g^{(k+2)}_{0A}=\langle g^{(l)}_{\mu\nu},l\leq k+1\rangle\ ,\ k\geq 0\ ,\  k\neq 1\ .
\end{align}
Thanks to (\ref{a3}), (\ref{a5}) and the following identity in dimension 2
\be\la{7a}
n_{AC}n^C_{\ B}=nn_{AB}+\frac 12(n_{CD}n^{CD}-n^2)s_{AB}
\ee 
equation (\ref{a4}) with $k=1$ reads   
\be\la{a6}
(p^B_{\ A}-\frac 14 nn^B_{\ A})_{|B}+\frac 18(n_{BC}n^{BC}- n^2)_{,A}=0\ .
\ee
Hence
\be\la{a7}
p_{AB}=\frac 18(n^2-n_{AB}n^{AB})s_{AB}-\frac 14nn_{AB}+\tilde p_{AB}\ ,
\ee
where $\tilde p_{AB}$ is a symmetric  TT-tensor 
\be\la{a8}
\tilde p^A_{\ A}=0\ ,\ \ \tilde p^{B}_{\ A|B}=0
\ee
on the sphere. In terms of the complex stereographic coordinates $\xi,\bar \xi$  solutions of (\ref{a8}) are given by $\tilde p_{AB}dx^Adx^B=Re(h(\xi)d\xi^2)$, where $h$ is a holomorphic function. The only regular solution  is $\tilde p_{AB}=0$. Due to this 
 one obtains
\be\la{a7a}
p_{AB}=\frac 18(n^2-n_{AB}n^{AB})s_{AB}-\frac 14nn_{AB}\ .
\ee

In order to analyse the remaining Einstein  equations we need a more explicit form of (\ref{14}). For $k\geq 1$ one obtains
\be\la{a7b}
Y^{(k)}+\frac{1}{k(k+1)}\tilde R_{11,0}^{(k-1)}=kg_{00}^{(k+2)}+\frac n2g_{00}^{(k+1)}+(g_{0A}^{(k+1)})^{|A}+\langle g^{(l)}_{\mu\nu},l\leq k\rangle\ ,\ k\geq 1\ .
\ee
Equation $\tilde R^{(k)}_{AB}=0$ with $k\geq 2$ can be splitted into its trace (with respect to $s_{AB}$) and a traceless part. The trace part 
\be\la{a10}
s^{AB}\tilde R^{(k)}_{AB}=0\ ,\ k\geq 2
\ee
 allows to obtain $g_{00}^{(k+2)}$ in terms of  lower order coefficients 
\be
g^{(k+2)}_{00}=\langle g^{(l)}_{\mu\nu},l\leq k+1\rangle\ ,\ k\geq 2\ .\la{a11}
\ee
The traceless part  is equivalent to the equation
\be\la{a13}
\tilde R^{(k)}_{AB}+\big(-\frac 12s^{CD}\tilde R^{(k)}_{CD}+\frac {1}{k+1}\tilde R_{11,0}^{(k-1)}\big)s_{AB}=0\ ,\ k\geq 2
\ee
which yields a simple differential condition for $g^{(k+1)}_{AB}$
\be\la{a14}
g^{(k+1)}_{AB,0}=\langle g_{0\mu}^{(k+1)},g^{(l)}_{\mu\nu},\ l\leq k\rangle\ ,\ k\geq 2\ .
\ee
Now equation $\tilde R_{11}^{(k-1)}=0$ plays a role of a constraint which is preserved by (\ref{a14}). Otherwise speaking, equation $\tilde R_{11}^{(k-1)}=0$  defines the trace of $g^{(k+1)}_{AB}$ with respect to $s^{AB}$, whereas (\ref{a14}) is an equation for the traceless part of $g^{(k+1)}_{AB}$.

In order to analyse equations $\tilde R_{AB}=0$  in the order $k=0,1$  we need  the following expansion 
of $\det{(g_{AB})}$ 
\be\la{a13a}
|g|=|\hat g|\big(1-n\Omega+(-p+\frac 12n^2-\frac 12n_{AB}n^{AB})\Omega^2\big)+O(\Omega^3)\ .
\ee
Equation $\tilde R_{AB}^{(0)} =0$  reads
\be
\hat R_{AB}-(b+\frac 12n_{,0})s_{AB}=0\ ,\la{21}
\ee
hence
\be\la{22}
b=1-\frac 12n_{,0}\ .
\ee
Using the standard identity in two dimenions
\be\la{22c}
R'_{AB}=\frac 12 R' g_{AB}\ ,
\ee
where $R'_{AB}$ is the Ricci tensor of $g_{AB}$, shows
that equation $\tilde R^{(1)}_{AB} =0$ does not carry any new information (it coincides with the $u$-derivative of (\ref{a7a})).

 The last equations to consider are $\tilde R^{(2)}_{00}=0$ and $\tilde R^{(2)}_{0A}=0$.
The first one   takes the form
 \be\la{22a}
M_{,0}=\langle n_{AB}\rangle\ .
\ee
This equation is responsible for diminishing of the gravitational energy if time $u$ increases.
Equation $\tilde R^{(2)}_{0A}=0$  yields
 \be\la{22b}
L_{A,0}=-\frac 13M_{,A}+\langle  n_{AB}\rangle\ .
\ee

We  summarize  consequences of the vacuum Einstein equations in the affine gauge in the following theorem.
\begin{thm}
Vacuum metric with smooth scri $\mathscr{I^+}$ can be transformed to the form
\be\la{3a}
\tilde g=du(\tilde g_{00}du+2dr+2\tilde g_{0A}dx^A)+\tilde g_{AB}dx^Adx^B\ ,
\ee 
\be\la{a10d}
\tilde g_{00}=1-\frac 12n_{,0}-\frac{2M}{r}+\Sigma_2^{\infty} g_{00}^{(k+2)}r^{-k}\ ,
\ee
\be\la{a10a}
\tilde g_{0A}=q_A+\frac{2L_A}{r} +\Sigma_2^{\infty} g_{0A}^{(k+2)}r^{-k}\ ,
\ee
\be\la{a12}
\tilde g_{AB}=-r^2s_{AB}+rn_{AB} +p_{AB}+\Sigma_3^{\infty} g_{AB}^{(k+2)}r^{-k}
\ee
with  coefficients  defined recursively in the following steps:
\begin{itemize}
 \item Tensor $n_{AB}$ can be arbitrary up to a gauge  condition e.g. $n=0$ (and unknown convergence conditions). It defines $q_A$ and $p_{AB}$  according to (\ref{a5}) and (\ref{a7a}). 
\item Coefficients $M$ and $L_A$ are  defined in quadratures 
by equation (\ref{22a}) and (\ref{22b}), respectively. 
\item
Trace $s^{AB}g_{AB}^{(3)}$ is given by (\ref{a2a}) with $k=1$ and 
the traceless part of $g_{AB}^{(3)}$ is defined in quadratures by equation  (\ref{a14}) with $k=2$.
\item
For $l\geq 4$ components $g_{00}^{(l)}$, $g_{0A}^{(l)}$ and $s^{AB}g_{AB}^{(l)}$ follow  directly from equations (\ref{a10}), (\ref{a4a}) and    (\ref{a2a}), respectively.
Then the traceless part of $g_{AB}^{(l)}$ is defined in quadratures by equation (\ref{a14}). 
\end{itemize}
\end{thm}
This analysis of the Einstein equations is equivalent to that in the luminosity gauge (see \cite{mw}).
The traceless part of $n_{AB,0}$  corresponds  to the Bondi news function.
Free data consist of 2 arbitrary functions on the boundary (the traceless part of $n_{AB}$) and initial values of $M$, $L_A$ and the traceless parts of $g_{AB}^{(l)}$ with $l\geq 3$ on a section  $u=u_0$ of the scri. The latter  fields are coefficients of an expansion of the traceless part of $\tilde g_{AB}$ restricted  to the 3-dimensional null surface $u=u_0$ approaching the scri. Unfortunately, we are not able to find conditions which guarantee convergence of series describing metric.

Above free data are given on the outgoing null surface and on the part of the null scri in the future of the surface. This combination differs from that assumed in mathematically sophisticated  existence theorems of  Kannar \cite{ka}, Chrusciel and Paetz \cite{cp} and others, where evolution of data is considered in the future of two intersecting null surfaces.(method of Rendall \cite{r}) or in the future of a lightcone (method of Dossa \cite{d}).
All these theorems are based on the Friedrich formulation of the conformal Einstein equations \cite{f1}. They assume free data  for functions different from those in the  Bondi-Sachs formulation. For above reasons they are not very helpful in solving the existence problem in our case.

In order to obtain the total energy at $u$=const one should pass from the affine gauge to the luminosity one. It means that we should replace coordinate $r$ by
\be\la{24a}
r_B=(\frac{\det\tilde g_{AB}}{\det s_{AB}})^{\frac 14}\ .
\ee
Practically it is sufficient  to consider an approximate formula
\be\la{28}
r_B=r-\frac 14n-\frac {1}{8r}n_{AB}n^{AB}+O(r^{-2})\ ,
\ee
which leads to the Bondi mass aspect
\be\la{34a}
M_B=M-\frac{1}{16}(n_{AB}n^{AB}-\frac{1}{2}n^2)_{,0}\ .
\ee
 The total energy is given by
\be\la{26}
E(u)=\frac{1}{4\pi}\int_{S_2}{M_Bd\sigma}\ .
\ee
Equation (\ref{22a}) assures that $E_{,0}\leq 0$ what is interpreted as a loss of energy due to emission of gravitational waves. 

\section{Stationary metrics}
If  metric admits  the smooth null scri and a timelike Killing vector $K$ then there are coordinates in which metric takes the form (\ref{3a}) with $u$-independent coefficients and $K=\p_u$. In order to show this let us first observe that any timelike vector must be null on the scri, so the Killing vector $K$ coincides with the null generator $\p_u$  on the scri. Let us fix
 a null surface $\Sigma_0$ intersecting $\mathscr{I^+}$ along a spherical surface $S_0$.  We endow $\Sigma_0$ into the following coordinates: $\Omega$ (affine distance from $S_0$ along null geodesics forming the surface) and $x^A$ (spherical coordinates transported from $S_0$ along the geodesics). Using the 1-parameter  group of motion $\phi_u$ related to $K$ we can generate from $\Sigma_0$ foliation of a neighbourhood $U$ by surfaces $u=const$. If we write the physical metric in  coordinates $(u,r=1/\Omega, x^A)$ and  transform $r$ appropriately we obtain (\ref{3a}) with coefficients independent of $u$.  
 The Killing field is given by $K=\p_u$ and we can choose the   conformal factor to be $\Omega=1/r$. The unphysical metric is given by (\ref{1}) and (\ref{2}) with $u$-independent coefficients. 

Let us consider low order Einstein equations (\ref{7}) in the stationary case. The regularity of the physical Ricci tensor $\tilde R_{\mu\nu}$ on $\mathscr{I^+}$ (see equations (\ref{10})-(\ref{11})) implies
 \be\la{62c}
g_{00}=b\Omega^2-2M\Omega^3+ O(\Omega^4)
\ee
and
\be\la{62d}
g_{0A}=q_A\Omega^2+2L_A\Omega^3+O(\Omega^4)\ .
\ee
In order to obtain $b=1$ we consider 
equation $\tilde R_{00}^{(2)}=0$. It  takes the form
\be\la{73}
\hat \Delta b=0\ ,
\ee
where $\hat\Delta$ is the Laplace operator related to the metric $\hat g_{AB}$ on the sphere. Multiplying (\ref{73}) by $b$ and integrating over the sphere shows that $b=$const. A rescaling of $u$ and $r$  allows to obtain 
\be\la{73a}
b=1\ .
\ee
Then equation $\tilde R_{AB}^{(0)} =0$  (see (\ref{21})) yields $\hat R=-2$ , hence
\be\la{74}
\hat g_{AB}=-s_{AB}\ .
\ee

Due to (\ref{73a}) and (\ref{74})  metric $\tilde g$ is  asymptotically Minkowskian in coordinates adapted to the Killing field $K$.  The remaining gauge freedom consists of   transformations of  coordinates $x^A$ preserving $s_{AB}$ (rotations of the sphere)
and supertranslations $u'\hat =u+f(x^A)$  combined with  a shift of the radial coordinate $r'=r+h(x^A)$. For  a later convenience we write  the latter two transformations in the linear approximation in $\Omega$ 
\be\la{63a}
u'=u+f-\frac 12 f^{,A}f_{,A}\Omega\ ,\ \ r'=r+h-f^{,A}(\frac 12f-h)_{,A}\Omega\ ,\ \ x'^A=x^A- f^{|A}\Omega\ .
\ee
They induce the following  transformation of $n_{AB}$ 
\be\la{63b}
n'_{AB}=n_{AB}+2f_{|AB}+2hs_{AB}\ .
\ee

Let us continue our analysis of the low order equations from the system (\ref{40a})-(\ref{40b}).  
Equation $\tilde R^{(2)}_{0A}=0$  reads
\be\la{75}
\frac 13M_{,A}=-q_{[B|A]}^{\ \ \ \ \ B}\ .
\ee
It can be written in the form
\be\la{75a}
\frac 13 dM=-{}^*d\alpha\ ,
\ee
where 
\be\la{75b}
\alpha=\eta^{AB}q_{A|B},
\ee
 $\eta_{AB}$ is the Levi-Civita tensor and star denotes the Hodge dual on the sphere. Since a differential form which is simultaneously exact and coexact on $S_2$  must vanish ione obtains
\be\la{77a}
M=const
\ee
and 
\be\la{77c}
\alpha=const\ .
\ee
Definition of $\alpha$ is equivalent to 
\be\la{76}
d(q_Adx^A)=-2\alpha\eta\ ,
\ee
where $\eta$ is the volume form corresponding to  $s_{AB}$. Due to (\ref{77c}) integration of (\ref{76}) over the sphere implies 
\be\la{76a}
\alpha=0\ .
\ee
Hence
\be\la{77}
q_A=q_{,A}\ ,
\ee
where $q$ is a function. Summarizing this part, equation $\tilde R^{(2)}_{0A}=0$ is equivalent to (\ref{77a}) and (\ref{77}).

Consider now equation  (\ref{a5}) in the gauge $n=0$ 
\be\la{78}
q_{,A}=\frac 12n^B_{\ A|B}\ .
\ee
In order to find consequences of (\ref{78}) we prove the following lemma.
\begin{lem}Every traceless tensor $n_{AB}$ on $S_2$ admits functions $F$, $H$ such that
\be\la{77b}
n_{AB}=F_{|AB}-\frac 12 \Delta Fs_{AB}+\eta^C_{\ (A}H_{|B)C}\ ,
\ee
where $\Delta$ is the standard Laplace operator on the sphere.
Functions $F$ and $H$ are defined up to $c^mY_{1m}+c$, where $c^m$ and $c$ are constants and $Y_{lm}$ are spherical harmonics.
\end{lem}
\noindent
\textbf{Proof.} Every tensor $n_{AB}$ defines the following form $\omega$ on $S_2$ 
\be\la{77e}
\omega=n_{A\ |B}^{\ B}dx^A\ ,
\ee
which can be decomposed into an exact and coexact form
\be\la{78c}
n_{A\ |B}^{\ B}=\tilde F_{,A}+  \eta^C_{\ A}\tilde H_{,C}\ .
\ee
We will show that given $\tilde F$ and $\tilde H$ there is solution $n'_{AB}$ of (\ref{78c}) of the form (\ref{77b}).

If $\tilde H=$const we assume 
\be\la{78d}
n'_{AB}=F_{|AB}-\frac 12 \Delta Fs_{AB}\ .
\ee
Then equation (\ref{78c}) yields
\be\la{78b}
(\Delta+2)F=2\tilde F+2c\ .
\ee
Expanding $F$ and $\tilde F$ into the spherical harmonics $Y_{lm}$ shows that only the dipole  part of $\tilde F$  has no counterimage in $F$. However, a direct analysis of  equation (\ref{78c}) with $\tilde F=c^mY_{1m}\neq 0$ and $\tilde H=$const shows that   regular  solutions $n_{AB}$ are not admitted in this case. It means that $\tilde F$ cannot contain harmonics $Y_{1m}$ and solution $F$ of  (\ref{78b}) always exists.  

If $\tilde F=$const  equation (\ref{78c}) can be written in the form
\be\la{80a}
{}^* n_{A\ |B}^{\ B}=\tilde H_{,A}
\ee
where ${}^*n_{AB}=\eta^{\ C}_{A}n_{CB}$ is also traceless and symmetric. 
Equation  (\ref{80a}) is satisfied by 
\be\la{80b}
{}^*n'_{AB}=H_{|AB}-\frac 12 \Delta Hs_{AB}
\ee
with $H$ satisfying
\be\la{80c}
(\Delta+2)H=2\tilde H+2c\ .
\ee
It follows from (\ref{80b}) that $n'_{AB}=\eta^C_{\ (A}H_{|B)C}$. 

Thus,  functions $\tilde F$ and $\tilde H$ defined by $n_{AB}$ can be also obtained from $n'_{AB}$ of the form (\ref{77b}). Tensor $n'_{AB}-n_{AB}$ is a (trivial) TT-tensor on $S_2$, so $n'_{AB}=n_{AB}$.
Note that solutions $F$ and $H$ of equations  (\ref{78b}) and (\ref{80c}) are defined  up to $c^mY_{1m}+c$.

$\Box$

\null

\noindent
Let us introduce the following symmetric, traceless operators on the sphere
\be\la{81b}
\nabla_{AB}=\nabla_A\nabla_B-\frac 12s_{AB}\bigtriangleup\ ,\ \  {}^*\nabla_{AB}=\eta_A^{\ C}\nabla_{CB}\ .
\ee
Due to Lemma  3.1  every symmetric traceless tensor on $S_2$ can be decomposed into linear combination of 
 \be\la{81c}
\nabla_{AB}Y_{lm}\ ,\ \ {}^*\nabla_{AB}Y_{lm}\ .
\ee
 Tensors (\ref{81c})  can be called tensor harmonics (compare with those in \cite{th} and references therein). In the same spirit every vector field on the sphere can be decomposed into
 \be\la{81d}
\nabla_{A}Y_{lm}\ ,\ \ {}^*\nabla_{A}Y_{lm}
\ee
which can be called vector harmonics.

Now we return to the Einstein equations.
It follows from (\ref{78}) and the proof of Lemma 3.1 that  $n_{AB}$ has  form (\ref{78d}).  Such $n_{AB}$ can be gauged away by means of transformation (\ref{63a}) with 
$f=-\frac 12F$ and $h=\frac 14\Delta F$.  Thus, there are coordinates adapted to the symmetry $\p_u$ such that
\be\la{81}
n_{AB}=0\ .
\ee
Now, it follows from (\ref{81}), (\ref{a7a}) and (\ref{78}) that
\be\la{81a}
p_{AB}=0\ ,\ \ q_A=0\ .
\ee
Moreover,  Lemma 3.1 shows that transformations (\ref{63a}) reduce  to the case 
\be\la{82}
f=c^mY_{1m}+c\ ,\ \ h= c^mY_{1m}\ .
\ee
Note that these residual transformations play  a role of the translation subgroup of the BMS group. Together with the group of rotations preserving $s_{AB}$ it forms  
the group of Euclidean motions of $R^3$  completed by time translations (constant $c$).

Equations $\tilde R_{AB}^{(k)}=0$ with $k=0,1$ are already exploited. For $k\geq 2$ in the nonstationary case  they were  used to define $g_{00}^{(k+2)}$ and $g_{AB,0}^{(k+1)}$. Since now $g_{AB,0}^{(k+1)}=0$
let us write  these equations in more detail
\begin{align}\la{91}
&[(k-1)g^{(k+2)}_{00}+(g_{0C}^{(k+1)})^{|C}-\frac 12 R'^{(k)}]s_{AB}-(k-1)g^{(k+1)}_{0(A|B)}\\\nonumber
&+ \frac{1}{2}k(k-1)g_{AB}^{(k)}=\langle g^{(l)}_{0\nu},\ g^{(l-1)}_{AB},l\leq k\rangle\ ,\ k\geq 2\ .
\end{align}
For $k=2$ the trace of (\ref{91}) yields 
\be\la{91a}
g_{00}^{(4)}=-L_C^{\ |C}\ ,
\ee
where $L_C=\frac 12g_{0C}^{(3)}$,
and the traceless part is
\be\la{69}
L_{(A|B)}=\alpha s_{AB}\ ,
\ee
where $\alpha$ is a function. It follows from ({\ref{69}) that $L^A\p_A$ is the conformal Killing field of the spherical metric. In terms of the position vector on the sphere $\bar r\in S_2\subset R^3$ the general smooth solution of (\ref{69})  is
\be\la{83}
L_Adx^A=\bar J(\bar r\times d\bar r)+\bar Dd\bar r\ ,
\ee
where $\bar J$ and $\bar D$ are constant vectors. Function $D=\bar D\bar r$ is a composition of harmonics $Y_{1m}$ so there is a chance to gauge it away by means of a transformation given by (\ref{63a}) and (\ref{82}). This transformation induces the following change
\be\la{84}
L'_A=L_A+Mf_{,A}\ .
\ee
Hence, for $M\neq 0$ we can eliminate $\bar D$ by taking $f=-\bar D\bar r/M$. Then we can use the rotation freedom to direct $\bar J$ along z-axis. Thus,  in the adapted  spherical coordinates $\theta$ and $\varphi$, for $M\neq 0$  one obtains 
\be\la{85}
L_Adx^A=J\sin^2{\theta}d\varphi\ 
\ee
An inspection of the Kerr metric shows  that constant $J$ is the total angular momentum. 

 In the  case $M=0$ we can use only the rotation group to reduce the number of free parameters in $\bar J,\bar D$ to 3. For instance one can obtain
\be\la{86}
L_Adx^A=J\sin^2{\theta}d\varphi+dD
\ee
where
\be\la{86c}
D=D_1\sin{\theta}\cos{\varphi}+D_2\sin{\theta}\sin{\varphi}+D_3\cos{\theta}
\ee
and either $D_1=0$ or $D_2=0$.  Here again $J$ is the total angular momentum (see a discussion after prolongation (\ref{117})-(\ref{120}) of metric to the spacelike infinity).

From  (\ref{85}) and (\ref{86}) one obtains
\be\la{86a}
L_C^{\ |C}=0\ \ if\ \ M\neq 0
\ee
\be\la{86b}
L_C^{\ |C}=-2D\ \ if\ \ M=0\ .
\ee
At this stage the physical metric is given by (\ref{3a}) with components
\be\la{87}
\tilde g_{00}=1-\frac{2M}{r}+\frac{2D}{r^2}+O(\frac {1}{r^3})\ ,\ \ M=const\ ,
\ee
\be\la{89}
\tilde g_{AB}=-r^2s_{AB}+O(\frac {1}{r})
\ee
and 
\be\la{88}
 \tilde g_{0A}dx^A=\frac {2J}{r} \sin^2{\theta}d\varphi+O(\frac {1}{r^2})\ ,\ \  D=0\ \ if\ M\neq 0
\ee
or
 \be\la{90}
\tilde g_{0A}dx^A=\frac 1r (2J\sin^2{\theta}d\varphi+2dD)\ ,\ if\ M=0
\ee
with $D$  given by (\ref{86c}).

Let us consider now equation (\ref{91}) with $k\geq 3$. Its trace defines  $g_{00}^{(k+2)}$ 
\be\la{100}
g^{(k+2)}_{00}= \frac{1}{(k-2)(k+1)}   \mathring g_{CD}^{(k)\ |CD}  + \langle g^{(k)}_{0\nu},g^{(l)}_{\mu\nu},\ l\leq k-1\rangle\ ,\ k\geq 3
\ee
and the traceless part reads
\be\la{93}
-g^{(k+1)}_{0(A|B)}+\frac 12 (g_{0C}^{(k+1)})^{|C}s_{AB}+\frac{1}{2}k\mathring g_{AB}^{(k)}=\langle g^{(k)}_{0\nu},g^{(l)}_{\mu\nu},\ l\leq k-1\rangle\ ,
\ee
where the mathring denotes the traceless part of a tensor,
\be\la{93b}
\mathring g_{AB}^{(k)}=g_{AB}^{(k)}-\frac 12(s^{CD}g_{CD}^{(k)})s_{AB}\ .
\ee
Still coefficients $g^{(k+1)}_{0A}$ can be expressed in terms of $g_{AB}^{(k)}$ due to (\ref{a4})
\be\la{93a}
g^{(k+1)}_{0A}=-\frac {k}{(k-2)(k+1)}g_{AC}^{(k)\ |C}+\langle g^{(l)}_{\mu\nu},l\leq k-1\rangle\ ,\ \ k\geq 3\ .
\ee
Substituting (\ref{93a}) into (\ref{93}) and eliminating $g_{0\mu}^{(k)}$  via (\ref{93a}) and (\ref{a11}) and trace of $g_{AB}^{(k)}$ via (\ref{a2a}) yields
\be\la{95}
-\mathring g_{C(A\ \ B)}^{(k)\ |C}+\frac 12\mathring g_{CD}^{(k)\ |CD}s_{AB}-\frac 12(k-2)(k+1)\mathring g_{AB}^{(k)}
=\langle g^{(l)}_{\mu\nu},l\leq k-1\rangle\ .
\ee
Equation (\ref{95}) can be written in a simpler form 
\be\la{96}
(\Delta+k^2-k-4)\mathring g_{AB}^{(k)}=\langle g^{(l)}_{\mu\nu},l\leq k-1\rangle
\ee
due to an  identity following from (\ref{22a})
\be\la{95a}
-\mathring g_{C(A\ \ B)}^{(k)\ |C}+\frac 12\mathring g_{CD}^{(k)\ |CD}s_{AB}+\frac 12\mathring g_{AB|C}^{(k)\ \ C}-\mathring g_{AB}^{(k)}
=\langle g^{(l)}_{\mu\nu},l\leq k-1\rangle\ ,
\ee
 however (\ref{95}) is better for further proceeding. Note that $\Delta$ in  (\ref{96}) is the covariant  Laplace operator on the sphere and it mixes indices in $\mathring g_{AB}^{(k)}$.

Using Lemma 2.1 let us represent tensor $\mathring g_{AB}^{(k+1)}$  by scalar potentials $Q^{(k)}$ and $P^{(k)}$ such that
\be\la{97}
\mathring g_{AB}^{(k+1)}=\nabla_{AB}Q^{(k)}+{}^*\nabla_{AB}P^{(k)}\ .
\ee
Equation (\ref{95}) splits into two  equations
\be\la{98}
(\Delta+k(k+1))Q^{(k)}=-2\tilde Q^{(k)}
\ee
\be\la{99}
(\Delta+k(k+1))Q^{(k)}=-2\tilde P^{(k)}\ ,
\ee
where  $\tilde Q^{(k)}$ and $\tilde P^{(k)}$ are potentials related to the r. h. s. of  (\ref{95}). A lenghty analysis of an explicit form of  equation (\ref{95}) shows that $\tilde Q^{(k)}$ and $\tilde P^{(k)}$
do not  contain $Y_{lm}$ with $l\geq k$. Hence, solutions $Q^{(k)}$ and $P^{(k)}$  exist and  are given up to $c^mY_{km}$.
These new parameters $c^m$ are then  implemented in $g_{0A}^{(k+2)}$ and $g_{00}^{(k+3)}$ via (\ref{93a}) and  (\ref{100}).

Below we summarize results of this section in terms of the physical metric $\tilde g$.
Note that the covariant derivatives, the Laplace operator $\Delta$, the Levi-Civita tensor $\eta_{AB}$ and the Hodge dual ${}^*$ are related to the spherical metric $s_{AB}$. 
\begin{thm}
Every stationary vacuum metric with a smooth conformal boundary $\mathscr{I^+}=R\times S_2$   can be transformed to 
the following form in a neighbourhood of $\mathscr{I^+}$ 
\be\la{99a}
\tilde g=du(\tilde g_{00}du+2dr+2\tilde g_{0A}dx^A)+\tilde g_{AB}dx^Adx^B\ ,
\ee 
where
\be\la{100}
\tilde g_{00}=1-\frac{2M}{r}+\frac{2D}{r^2}+\Sigma_{k=2}^{\infty}\frac{k(k+1)}{r^{k+1}}(Q^{(k)}+l.o.)\ ,
\ee
\be\la{102}
\tilde g_{0A}dx^A=\frac 1r( 2J\sin^2{\theta}d\varphi+2dD)+\Sigma_{k=2}^{\infty}\frac {k+1}{r^k}[d(Q^{(k)}+l.o.)+{}^*d(P^{(k)}+l.o.)],
\ee
\be\la{103}
\tilde g_{AB}=-r^2s_{AB}+\Sigma_{k=2}^{\infty}\frac {1}{r^{k-1}}[\nabla_{AB}(Q^{(k)}+l.o.)+{}^*\nabla_{AB}(P^{(k)}+l.o.)+(l.o.)s_{AB}]\ ,
\ee
\be\la{104}
Q^{(k)}=\Sigma_{m=-k}^{m=k}Q^{km}Y_{km}\ ,\ \ P^{(k)}=\Sigma_{m=-k}^{m=k}P^{km}Y_{km}\ .
\ee
and $l.o.$ (lower order) denotes terms spanned by $Y_{lm}$ with $l<k$ and depending  on $M$, $a$, $Q^{lm}$ and $P^{lm}$ with $l<k$ (no such terms for $k=2$). If $M= 0$ then  $D$ is given by (\ref{86c}), otherwise $D=0$. 
\end{thm}

\noindent
\textbf{Remark.} The assumption of smoothness of the scri in Theorem 3.1 may be replaced by the assumption that $\Omega$, $\tilde g(K)$ and $g$ are of the class $C^2$ in a neighbourhood $U$ of $\mathscr{I^+}$. Indeed, under these assumptions  all considerations in this section up to equations (\ref{87})-(\ref{90}) are still valid.  We can  introduce coordinate $t$ via $u=t-r$ and continue metric analytically in $t$ to all values of $t$. For large values of $r$ surfaces $t=const$ are spacelike and metric $\tilde g$ is stationary and asymptotically flat at spacelike infinity. A generalization of results of Beig and Simon \cite{bs} by Kundu \cite{k}   assures analyticity of the Ernst potential and compactified 3-dimensional metric  in harmonic coordinates if $M\neq 0$. Then one can introduce spherical coordinates based on the normal coordinates at point at infinity.   Arguments of  Damour and Schmidt (see Appendix  in \cite{ds}) show that 4-dimensional spacetime metric $\tilde g$ should  admit an analytic compactification  up to the scri $\mathscr{I^+}$. Knowing that $\Omega$ and $g$ are analytic   we can construct analytic foliation of $U$ and the Bondi-Sachs coordinates along lines in the begining of this section. Then all components of $g$ should be analytic functions.

\null

Constants $Q^{km}$ and $P^{km}$ are multipole moments of metric.  The easiest way to find $Q^{km}$ is to integrate (\ref{100}) with spherical harmonics. Moments $P^{km}$ arise if expression  (\ref{102}) is  integrated with $dY_{lm}$.
For  each $k\geq 2$ up to  $4k+2$ multipole moments  are admitted.  
They are restricted  by yet unknown  convergence conditions.
In the axially symmetric case only 2 parameters for each $k$ can appear ($Q^{k0}$ and $P^{k0}$).

In order to identify $Q^{km}$ and $P^{km}$ with multipole moments introduced by Thorne \cite{th} let us replace coordinate $u$ by $t$ defined by
\be\la{114}
u=t-\int{\frac{dr}{\tilde g_{00}}}\ .
\ee
Note that $t$ is given up to a function on the sphere.   For some choice of this function expansion of (\ref{114}) takes the form
\be\la{115}
u=t-r-2M\ln{(\frac{r}{2M}-1)}-\frac{2D}{r}-\Sigma_{k=2}^{\infty}\frac{(k+1)}{r^k}(Q^{(k)}+l.o.)\  .
\ee
Let $\tilde g'_{\mu\nu}$ denote components of metric in coordinates $t,r,x^A$.
Substituting (\ref{115}) into  equations (\ref{100})-(\ref{103}), or their counterparts for $M=0$, yields
\be\la{117}
\tilde g=\tilde g_{00}dt^2+2\tilde g'_{0A}dtdx^A-\frac{dr^2}{\tilde g_{00}}+2\tilde g'_{1A}drdx^A+\tilde g'_{AB}dx^Adx^B\ ,
\ee 
where
\be\la{118}
\tilde g'_{0A}dx^A=\frac {2J}{r} \sin^2{\theta}d\varphi+\Sigma_{k=2}^{\infty}\frac {k+1}{r^k}[d(l.o.)+{}^*d(P^{(k)}+l.o.)]\ ,
\ee
\be\la{119}
\tilde g'_{1A}dx^A=-\tilde g_{0A}dx^A+\Sigma_{k=2}^{\infty}\frac {1}{r^k}[d(l.o.)+{}^*d(l.o.)]\ ,
\ee
\be\la{120}
\tilde g'_{AB}=\tilde g_{AB}+\Sigma_{k=3}^{\infty}\frac {1}{r^{k-1}}[d(l.o.)+{}^*d(l.o.)]
\ee
and $l.o.$ means harmonics of order smaller than $k$. It follows from (\ref{117})-(\ref{120}) that $t$ is a timelike coordinate for sufficiently big $r$. The exterior curvature form $K_{ij}$ of surface $t=const$ satisfies
\be\la{120a}
K_{11}=0(r^{-5})\ ,\ \ K_{1A}dx^A=\frac 1rJ\sin^2{\theta}d\varphi\ ,\ \ K_{AB}=0(r^{-1})\ .
\ee 
The ADM formula for the linear momentum $P^i$ shows that $P^i=0$. Using Proposition 2.2 in \cite{chwy} allows to identify $J$ as the total angular momentum. 

For $M\neq 0$  formulas (\ref{117})-(\ref{120}) have the form (11.4) in \cite{th}  (note that $Y_j^{B,lm}dx^j$ in \cite{th} coincides with ${}^*dY^{lm}$), hence 
\be\la{121}
l(l+1)Q^{lm}=\frac 12(2l-1)!! \big(\frac{2l(l-1)}{(l+1)(l+2)}\big)^{1/2}I^{lm}\  ,
\ee
\be\la{122}
(l+1)P^{lm}=-\frac 12(2l-1)!! \big(\frac{2(l-1)}{l+2}\big)^{1/2}S^{lm}\ ,
\ee
where $I^{km}$ and  $S^{km}$ are multipole moments of Thorne. For $M=0$ our expression for the square of the Killing vector $\tilde g_{00}=K^2$ contains the dipole mass moment $D$ (this term is assumed to be constant in \cite{th}). This is not in contradiction to \cite{th}  since the case $M=0$ is not considered there.

Since  $I^{lm}$, $S^{lm}$ are equivalent to the STF moments of Thorne (see equations (11.2)-(11.3) in \cite{th}) and the latter were proved  \cite{gu} to be equivalent to moments  of Geroch \cite{g} and Hansen \cite{h} all properties related to  these moments can be  translated into our formalism. The only  exception is the case $M=0$.  As we have shown then the nontrivial mass dipole moment $D$ is admitted. The known statement about asymptotic Kerr like behaviour of stationary metrics (see \cite{bs1,th} and references therein) should be clarified. We can rephrase this statement in the following way:

\begin{prop} 
	Every asymptotically flat stationary vacuum metric with $M\neq 0$ or $K^2=1+0(r^{-3})$
	  tends to the Kerr metric in the order $r^{-2}$ with repect to asymptotically Minkowskian coordinates.
\end{prop}

Below we show how to compute the Bondi-Sachs coordinates for the Kerr metric with accuracy sufficient to find 
the quadrupole moments. We start with the standard Boyer-Lindquist coordinates. In order to define foliation  $u=const$ we  consider generalization of   the Eddington-Finkelstein retarded time. A correction of the order $O(1)$ is removable by a supertranslation, so as the function $u$  we take
\be\la {105}
u=t-r-2M\ln{(\frac{r}{2M}-1)}-\frac{A(\theta)}{r}+O(\frac{1}{r^2})\ .
\ee
Condition $u^{,\alpha}u_{,\alpha}=0$ implies
\be\la{106}
A=\frac 12a^2\sin^2{\theta}\ .
\ee
New coordinates $r',\theta',\varphi'$ should satisfy $u^{,\alpha}r'_{,\alpha}=1$, $u^{,\alpha}\theta'_{,\alpha}=0$, $u^{,\alpha}\varphi'_{,\alpha}=0$. Assuming that they are given by  $r,\theta,\varphi$ plus corrections one obtains
\be\la{107}
r'=r+\frac Ar+O(\frac{1}{r^2})\ ,\ \ \theta'=\theta+O(\frac{1}{r^2})\ ,\ \ \varphi'=\varphi+O(\frac{1}{r^2})\ .
\ee
 Inverting relations (\ref{105}) and (\ref{107}) yields
 \be\la {108}
t=u+r'+2M\ln{(\frac{r'}{2M}-1)}+O(\frac{1}{r'^2})\ ,\ \ r=r'-\frac {a^2\sin^2{\theta'}}{2r'}+O(\frac{1}{r'^2})\ ,
\ee
\be\la{109}
\theta=\theta'+O(\frac{1}{r'^2})\ ,\ \ \varphi=\varphi'+O(\frac{1}{r'^2})\ .
\ee
Now we substitute (\ref{108}) and (\ref{109}) into the Kerr metric. It is easy to check that there is no term in $\tilde g'_{AB}$ linear in $r'$. Thus, condition (\ref{81}) is satisfied in the new coordinates. The only term of the order $1/r^2$ in $\tilde g'_{0A}dx^A$ follows from $dt^2$. Hence it must be an exact form, so
\be\la{111}
P^{(2)}=0\ .
\ee 
Component $Q^{(2)}$ can be aesily 
 computed from the original component $g_{00}=K^2$  of the Kerr metric. Hence
\be\la{110}
Q^{(2)}=\frac 16 Ma^2(3\cos^2{\theta}-1)
\ee
and
\be\la{112}
Q^{20}=\frac 23\sqrt{\frac{\pi}{5}}Ma^2\ ,\ \ Q^{2A}=P^{20}=P^{2A}=0\ .
\ee
Thanks to Theorem 3.1 we can write the quadrupol approximation of the Kerr metric in the Bondi-Sachs coordinates without knowledge of further terms in transformations (\ref{105}) and (\ref{107})
\begin{align}\la{113}
&du[(1-\frac{2M}{r'}+\frac{Ma^2}{2r'^3}(\cos^2{\theta'}-\frac 13))du+2dr'+\frac{4Ma}{r'}\sin^2{\theta'}d\varphi'-\frac{3Ma^2}{2r'^2}\sin{2\theta'}d\theta']\nonumber\\
&-(r'^2+\frac{Ma^2}{2r'}\sin^2{\theta'})(d\theta'^2+\sin^2{\theta'}d\varphi'^2)+\frac{Ma^2}{r'}\sin^2{\theta'}d\theta'^2\ .
\end{align}

\section{Summary}
In section 2 we investigated the vacuum Einstein equations for metrics admitting the smooth null scri $\mathscr{I^+}$. We wrote these metrics in the Bondi-Sachs form using  the affine gauge ($\tilde g_{01}=1$) instead of the  luminosity gauge (\ref{24a}). We defined a minimal set of independent equations (\ref{40a}) and (\ref{40b}). Expanding  metrics and equations into powers of $1/r$ led us to a hierarchy of equations which can be solved recursively (see Theorem 2.1). This is an approach parallel to the standard one using  the luminosity gauge. The problem of  convergence of resulting series is still unsolved.

In section 3 we assumed that metrics admit additionally  a timelike Killing vector $\p_u$. Using the low order Einstein equations we showed that the approach from section 2 is still aplicable with omitted dependence on $u$.
The main results  are given in Theorem 3.1 describing an asymptotic form of metric. In agreement with the asymptotic analysis at spacelike infinity every solution with nonvanishing mass $M$  tends to the Kerr metric. Multipole moments  of two kinds appear  in consecutive orders of $1/r$.   A relation between them and one set of moments of Thorne is given by (\ref{121}) and (\ref{122}). Our results on metrics with $M=0$ led us to a slight clarification (see Proposition 3.1)  of statement about the Kerr like behaviour of stationary metrics.
 In the last part of  section 3 we found the approximate Bondi-Sachs coordinates for the Kerr metric and we wrote this metric up to first terms including the quadrupole moments.

\null

\noindent
\textbf{Acknowledgments}

\null

\noindent
I am grateful to Piotr Chru\'sciel and Walter Simon for useful discussions and pointing out references \cite{bs,ds}. This work was partially   supported by Project OPUS 2017/27/B/ST2/02806
of Polish National Science Centre (NCN).


\begin{thebibliography}{99}
\bibitem{ac}
Andersson L and  Chrusciel P T 1993 Hyperboloidal Cauchy data for vacuum Einstein equations and obstructions to smoothness of null infinity
\textit{Phys. Rev. Lett.} \textbf{70} 2829-2832 
\bibitem{acf}
Andersson L,  Chru´sciel P T and Friedrich H 1992 On the Regularity of Solutions of the Yamabe Equation and the Existence of Smooth Hyperboloidal Initial Data for Einstein Field
Equations \textit{Commun. Math. Phys.} \textbf{149} 587 

\bibitem{bs}
Beig B and Simon W 1980 Proof of a Multipole Conjecture due to Geroch \textit{Commun. Math. Phys.} \textbf{78} 75-82

\bibitem{bs1}
Beig B and Simon W 1980
The Stationary Gravitational Field Near Spatial Infinity \textit{Gen. Rel. Grav.} \textbf{12} 1003

\bibitem{b}
Bondi H 1960  Gravitational Waves in General Relativity \textit{Nature} \textbf{186} 535

\bibitem{bbm}
Bondi H, van der Burg M G J and Metzner A W K 1962 Gravitational Waves in General Relativity. VII. Waves from Axi-Symmetric Isolated Systems.
\textit{Proc. R. Soc. London}  \textbf{A 269} 21

\bibitem{chwy}
Po-Ning Chen, Lan-Hsuan Huang, Mu-Tao Wang and Shing-Tung Yau 2014 On the Validity of the Definition of Angular Momentum in General Relativity \textit{Annales Henri Poincare} \textbf{17}(2)

\bibitem{cms}
Chrusciel P T,  MacCallum M A H and Singleton D 1995 Gravitational Waves in General Relativity. XIV: Bondi Expansions and the ``Polyhomogeneity'' of Scri
\textit{Phil. Trans. Royal Soc. of London} \textbf{A350} 113-141 

\bibitem{cp}
Chrusciel P T and Paetz T-T 2013
Solutions of the vacuum Einstein equations with initial data on past null infinity  \textit{Class. Quantum Grav.} \textbf{30}  235037

\bibitem{ds}
Damour T and Schmidt B 1990 Reliability of perturbation theory in
general relativity
\textit{J. Math. Phys} \textbf{31}  2441 

\bibitem{d}
Dossa M 2003 Probl\`emes de Cauchy sur un Conoide Caracteristique pour les Equations d’Einstein (Conformes) du Vide et pour les Equations de Yang-Mills-Higgs \textit{Ann. Henri Poincar\'e} \textbf{4}  385 – 411


\bibitem{f1}
Friedrich H 1981
On the Regular and the Asymptotic Characteristic Initial Value Problem for Einstein's Vacuum Field Equations \textit{Proc. R. Soc. Lond.} \textbf{A 375} 169-184 

\bibitem{f}
Friedrich H 1983     Cauchy problems for the conformal vacuum field equations in general relativity
Commun. Math. Phys. \textbf{91} 445–472

\bibitem{g}
Geroch R 1970 Multipole Moments. II. Curved Space \textit{J. Math. Phys.} \textbf{11} 2580 

\bibitem{gu}
Gürsel Y 1983 Multipole moments for stationary systems: The equivalence of the Geroch-Hansen formulation and the Thorne formulation. \textit{Gen. Relat. Gravit.} \textbf{15} 737–754 

\bibitem{h}
Hansen R O 1974 
Multipole moments of stationary space-times \textit{J. Math. Phys.} \textbf{15} 46

\bibitem{ka}
Kannar J 1996 
On the Existence of $C^{\infty}$ Solutions to the Asymptotic Characteristic Initial ValueProblem in General Relativity \textit{\textit{Proceedings: Mathematical, Physical and Engineering Sciences}} \textbf{452} (No 1947)  945-952

\bibitem{k}
Kundu P 1981 On the analyticity of stationary gravitational
fields at spatial infinity
\textit{J. Math. Phys} \textbf{22} 2006 

\bibitem{mw}
 Mädler T and  Winicour J 2016 Bondi-Sachs Formalism, \textit{Scholarpedia} \textbf{11}(12):33528
\bibitem{p}
Penrose R 1963 Asymptotic Properties of Fields and Space-Times \textit{Phys. Rev. Lett.} \textbf{10} 66

\bibitem{r}
Rendall A D 1990  Reduction of the characteristic initial value problem tothe Cauchy problem and its applications to the Einstein equations  \textit{Proc. Roy. Soc. London} \textbf{A 427 }221–239

\bibitem{s}
Sachs R 1962 Gravitational Waves in General Relativity. VIII. Waves in Asymptotically Flat Space-Time.
\textit{Proc. R. Soc. London}  \textbf{A 270} 103

\bibitem{th}
Thorne Kip S 1980 
Multipole expansions of gravitational radiation \textit{Rev. Mod. Phys.} \textbf{52} 299 

\bibitem{t1}
Trautman A 1958 Radiation and Boundary Conditions in the Theory of Gravitation \textit{Bull. Acad. Pol. Sci., Ser. Sci. Math. Astron. Phys.}  \textbf{6} 407-412
\bibitem{t2}
Trautman A 1958 Lectures on General Relativity, mimeographed notes, King’s College, reprinted 2002  \textit{Gen. Relat. Grav.} \textbf{34} 721–762

\bibitem{w}
Winicour J 1985 Logarithmic asymptotic flatness \textit{Found. Phys.} \textbf{15}  605–616 


\bibitem{w1}
Winicour J 2012 Characteristic Evolution and Matching   \textit{Living Rev. Relativity} \textbf{15} 2 
\end{thebibliography}
\end{document}